\documentclass[aps,pra,twocolumn,10pt,showpacs]{revtex4-1}

\usepackage{amsmath}
\usepackage{tabularx}
\usepackage{graphicx}
\usepackage{times}
\usepackage{amssymb}
\usepackage{siunitx}
\usepackage{hyperref}
\usepackage{textcomp}
\usepackage{xcolor}
\usepackage[T1]{fontenc}
\usepackage{ulem}
\usepackage{url}
\usepackage{pdfpages}
\newcommand{\dyad}[2]{%
	| #1\left\rangle \right\langle  #2 |
}
\newcommand{\dif}[2]{%
\frac{\mathrm{d}}{\mathrm{d}#1} #2
}

\newcommand{\mean}[1]{%
\left\langle #1 \right\rangle
}

\begin{document}
\title{Nonclassical light from an incoherently pumped quantum dot in a microcavity}
\author{L. Teuber}
\email{Electronic address: lucas.teuber@uni-rostock.de}
\author{P. Gr\"unwald}
\author{W. Vogel}
\affiliation{Institut f\"ur Physik, Universit\"at Rostock, D-18055 Rostock, Germany}
\date{\today} 

\begin{abstract}
Semiconductor microcavities with artificial single-photon emitters have become one of the backbones of semiconductor quantum optics. In many cases, however, technical and physical issues limit the study of optical fields to  incoherently excited systems. We analyze the model of an incoherently driven two-level system in a single-mode cavity. The specific structure of the applied master equation yields a recurrence relation for the steady-state values of correlations of the intracavity field and the emitter. We provide boundary conditions that permit a systematic solution which is numerically less demanding than standard methods. The method allows us to directly infer reasonable cutoff conditions from the system parameters. Different cavity systems from previous experiments are analyzed in terms of field correlation functions which can be measured via homodyne correlation measurements. We find that nonclassical correlations occur in systems of moderate quantum-dot--cavity coupling rather than strong coupling. Our boundary conditions also allow us to derive analytical results for the overall quantum state and its higher-order moments. We obtain very good approximations for the full quantum state of the field in terms of the characteristic functions.It turns out that for every physically reasonable set of system parameters the state of the intracavity field is nonclassical.
\end{abstract}


\maketitle

\section{Introduction}
The basic structure of a two-level system (TLS), located in a quasi-resonant single-mode cavity, shows remarkable quantum properties. Already in the regime of weak coupling between the intracavity field and the TLS, the emission rate of the latter is increased by the so-called Purcell factor~\cite{P46}. For stronger coupling, the energy eigenvalues of the system are drastically changed, resulting in dressed states (polaritons) and Rabi splittings~\cite{Rempe92}, which are also known from strong atom-laser interaction~\cite{Mollow,Stroud}. The mentioned phenomena are pure quantum effects indicating the quantum nature of the systems under study. Thus, such a setup also constitutes a versatile source for nonclassical light. Antibunching and sub-Poisson photon statistics were shown in experiments with ions in optical cavities~\cite{walther,toschek}. For details, see also~\cite{C99,C08}.

More recently, semiconductor microcavities became a focus of research. In these systems, excitons in quantum dots adopt the role of the TLS~\cite{ReiFor10}. Semiconductor microcavities are much more complex than their atomic counterparts, as both the cavity and the quantum dot are embedded in an interacting medium. In particular, the dissipation rates of dot and cavity are usually dominant, limiting the possibility of strong coupling. Nevertheless, antibunching and sub-Poisson photon statistics could be demonstrated for quantum dots themselves~\cite{MichlerExp,Shih-2} as well as inside microcavities~\cite{Shih-1}. Furthermore, Rabi splitting was achieved in some realizations of semiconductor microcavities~\cite{QDstrong1,QDstrong2,QDstrong4}.

Another peculiar aspect in semiconductor microcavities is the difficulty to drive them with coherent light. This is due to multiple factors such as intensive light scattering off the sample geometry, resonance frequencies unfavorable for current laser technologies, and others~\cite{Shih-2}. As a consequence, incoherent excitation is a broadly applied method in semiconductor optics, e.g., via far-detuned photoluminescence~\cite{QDstrong1} or electroluminescence~\cite{Xuan11}. Theoretical descriptions of the quantum fields emitted from such systems are of great relevance to semiconductor quantum optics. For excitons behaving like bosonic particles, it has been shown that the regimes of weak and strong coupling depend sensitively on the system parameters~\cite{QDstrong3}.

An early treatment of the steady-state properties of incoherently driven TLSs in cavities was given by Agarwal and Dutta Guppta~\cite{AD90}. They applied continued fraction methods to obtain solutions and compared them to full numerical calculations. Nonclassical phenomena of the radiation field had been studied on the basis of first- and second-order moments of the photon number statistic by inspection of Mandel's $Q$ parameter~\cite{Ma79}. This is a first important step, which gives insight in the sub-Poisson photon statistics. Later on, more general criteria for the nonclassicality of light, such as matrices of moments of the photon number operator~\cite{AgTa}, had been introduced. This method was further generalized to a full characterization of the quantum properties of light through higher-order moments of two noncommuting observables~\cite{SRV05,EvgenNC05}. Alternatively, general nonclassicality tests can be based on the characteristic function of the radiation field~\cite{V00,RV02}. Such methods have not been applied yet to the systems under study.

In the present paper we study the quantum properties of an incoherently pumped TLS in a microcavity. Due to the incoherent dynamics, only specific correlations couple with each other. This yields closed infinite sets of coupled equations. In the steady-state case, correlations between the quantum dot and the intracavity field follow from recurrence relations, which are numerically less demanding than previous methods. We will solve these equations by applying boundary conditions following directly from the recurrence relations. The boundary conditions not only yield the necessary criteria for the solution but also an analytical proof of convergence. The structure of the solution allows us to determine an appropriate cutoff for numerical calculations, which solely depends on the system parameters. The boundary conditions directly imply properties of the quantum state of light. In particular, the state of the intracavity field can never resemble a thermal one under realistic physical conditions. Based on general moments criteria for nonclassicality, we analyze various quantum effects of the system for experimental parameters of different cavities studied within the last decade. It turns out that the system with moderate quantum-do--cavity coupling shows stronger signatures of nonclassicality than for stronger coupling. Finally, we apply the boundary conditions to approximate the characteristic function of the quantum state of the intracavity field with controlled errors. In this way we prove the nonclassicality of the latter for any incoherently pumped quantum-dot--cavity system.

The article is structured as follows. In Sec.~\ref{sec:system}, we specify the system under study and give the corresponding equations of motion. Section~\ref{sec:ansatz} is used to introduce our ansatz for solving the steady-state case using recurrence techniques. We also derive the above mentioned boundary and initial conditions there. In Sec.~\ref{sec:nonclass}, we test various nonclassicality criteria with parameters describing realistic systems applied in recent years.
Section~\ref{sec:charfct} deals with the characteristic function of the intracavity field, including a general treatment of the asymptotic behavior of the recurrence relation.
In Sec.~\ref{sec:sum}, we give a summary and some conclusions.

\section{System}\label{sec:system}
\begin{figure}[h]

\centering

\includegraphics[width=8cm]{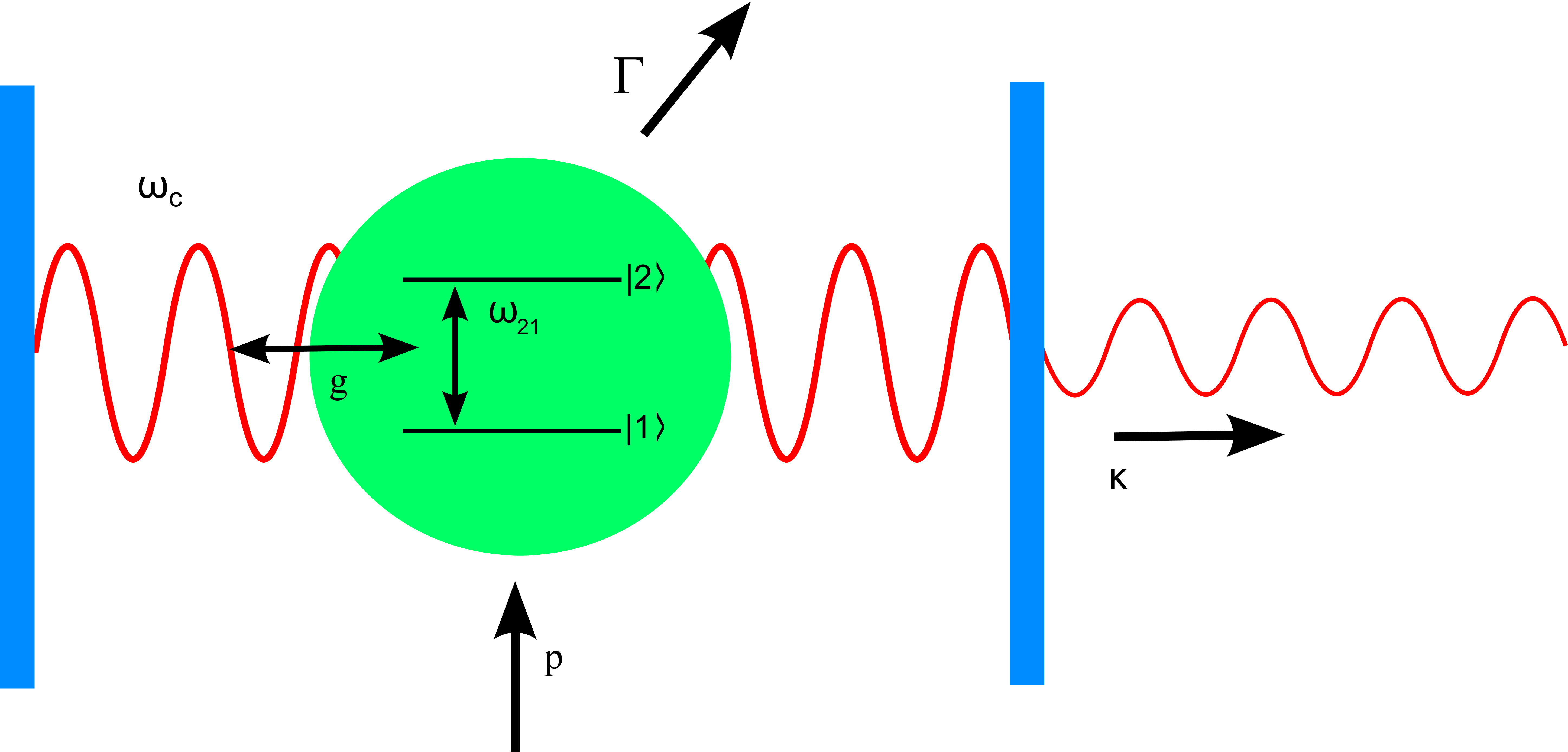}

\caption{(Color online) Scheme of a quantum dot (two-level system) in a (semiconductor) microcavity. The left mirror is supposed to reflect perfectly, whereas the right one is partially transparent giving rise to a loss of the cavity mode.}

\label{fig:scheme}

\end{figure}

A scheme of the model under study with the considered processes is depicted in Fig.~\ref{fig:scheme}.
The quantum dot in a microcavity is described by a TLS with transition frequency $\omega_{21}$ coupled to a mode of the electromagnetic field inside the cavity with frequency $\omega_{c}=\omega_{21}+\delta$. Applying the dipole approximation and the rotating wave approximation, and shifting into the frame rotating with $\omega_{21}$, we obtain the Jaynes-Cummings Hamiltonian~\cite{JC63}
\begin{align}
\label{eq:hamilton}
\hat{H} =  \hbar\delta \, \hat{a}^\dagger\hat{a} + \hbar g \left( \hat{a}^\dagger\hat{A}_{12} + \hat{A}_{21}\hat{a} \right).
\end{align}
Here, $\hat{a}$ and $\hat{a}^{\dagger}$ are the annihilation and creation operator of the field mode, respectively, $\hat{A}_{nm}=\dyad{n}{m}$ ($m,n=1,2$) are the atomic operators of the TLS, and $g$ denotes the coupling strength.

The dynamics of the system under incoherent pumping is governed by a master equation for the system density operator $\hat{\rho}$~\cite{C08}, which reads as
\begin{align}
\label{eq:master}
\dif{t}{\hat{\rho}} = - \frac{i}{\hbar} \left[ \hat{H} , \hat{\rho}\right] + \frac{\Gamma}{2} \mathcal{L}_{\hat{A}_{12}}(\hat{\rho}) + \frac{p}{2} \mathcal{L}_{\hat{A}_{21}}(\hat{\rho}) + \frac{\kappa}{2} \mathcal{L}_{\hat{a}}(\hat{\rho}),
\end{align}
with pumping strength $p$, quantum-dot spontaneous-emission rate $\Gamma$, and cavity decay rate $\kappa$. The relaxation processes are incorporated via the Lindblad operators $\mathcal{L}_{\hat{X}}(\hat{\rho})~=~[ \hat{X}\hat{\rho}, \hat{X}^{\dagger} ]+[ \hat{X},\hat{\rho}\hat{X}^{\dagger} ]$.

In the following, we write down all correlations in normal ordered form for the sake of clarity.
The structure of Eq.~\eqref{eq:master} has the effect that only those moments $\mean{\hat{A}_{21}^{k}\hat{A}_{12}^{l}\hat{a}^{\dagger \, m}\hat{a}^{n}}$ ($k,l,m,n\in\mathbb N$) that create or annihilate the same number of excitations overall (that is, both intracavity field and TLS) couple to each other.
Furthermore, due to the incoherent nature of the pumping process, those moments with excess creation or annihilation vanish in the long-time limit as they do not couple to the only inhomogeneous correlation $1=\langle\hat{A}_{21}^{0}\hat{A}_{12}^{0}\hat{a}^{\dagger \, 0}\hat{a}^0\rangle $. There remain three types of independent, nonvanishing moments, which we abbreviate as
\begin{subequations}\label{eq:momentsdef}
\begin{align}
I_n &= \mean{\hat{a}^{\dagger\,n}\hat{a}^n}, \\
B_n &= \mean{\hat{A}_{22}\hat{a}^{\dagger\,n}\hat{a}^{n}},\\
R_n &= \mean{\hat{A}_{21}\hat{a}^{\dagger\,n}\hat{a}^{n+1}}.
\end{align}
\end{subequations}
The corresponding set of coupled equations of motion can easily be derived from~\eqref{eq:master} and reads
\begin{subequations}
\label{eq:DES1_sub}
\begin{align}
\dif{t}{I_n}&=-n\kappa I_n -2ng \, \mathrm{Im}[ R_{n-1}],\\ 
\dif{t}{B_n}&= - \sigma_{n} B_n + p I_n + 2g \, \mathrm{Im}[R_n], \\
\label{eq:R_n}
\dif{t}{R_n}&=-\left( i\delta + \gamma_{n} \right) R_n \notag \\ & \qquad-ig [(n+1)B_n + 2\, B_{n+1} - I_{n+1} ],
\end{align}
\end{subequations}
where we defined $\gamma_n=\left(\Gamma+p+\kappa(2n+1)\right)/2$ and $\sigma_n=\Gamma+p+n\kappa$.
On a side note, we state that for certain limiting conditions, analytical solutions can be found; e.g., in the steady state for $\kappa \to 0$, we find a thermal state for the intracavity moments $I_n$. In the following, we consider all dissipation rates and $g$ to be finite but nonzero, and define this as a physically reasonable set of system parameters.

It should be noted that, in general, semiconductor quantum dots also experience strong nonradiative dephasing. This can be easily included by adding a Lindblad term $\tfrac{\Gamma_\text D}{2}\mathcal L_{\hat A_{22}}(\hat\rho)$. It couples only to the moments $R_n$ yielding a different value for $\gamma_n$, which reads
\begin{equation}
  \tilde\gamma_n=\gamma_n+\frac{\Gamma_\text D}{2}.
\end{equation}
For our purpose, it does not change any of our qualitative statements and the amplitude of the considered correlations varies only slightly. Furthermore, for the realistic cavities considered in Secs.~\ref{sec:nonclass} and~\ref{sec:charfct} no values for $\Gamma_\text D$ were given. Hence, we discard these dephasing effects throughout the paper.

\section{Ansatz for steady state solutions}\label{sec:ansatz}
We are interested in the long-time solution (steady state) of the set of equations~\eqref{eq:DES1_sub}. In this case, Eq.~\eqref{eq:DES1_sub} reduces to a homogeneous and coupled set of algebraic equations. The specific structure of these equations makes it feasible to formulate a recurrence relation for the moments $I_{n}$ of the form
\begin{align}
\label{eq:rec1}
I_{n+2} = \alpha_{n+1} I_{n+1} + \beta_n I_n,
\end{align}
with the recurrence coefficients $\alpha_{n}$, $\beta_{n}$ given by
\begin{align}\label{eq:alpha}
\alpha_n &= \frac{\sigma_n}{2 \kappa} \left( \frac{2 p}{\sigma_n} - \frac{n \kappa}{\sigma_{n-1}} - \frac{1 + \Lambda_{n-1}}{\Lambda_{n-1}} \right),\\
\label{eq:beta} \beta_n &= \frac{(n+1)\, p}{2 \kappa} \frac{\sigma_{n+1}}{\sigma_n},
\end{align}
where we set $\Lambda_n=\left( 2g^2 \gamma_n \right) / \kappa \left( \delta^2 + \gamma_{n}^2 \right)$. 

In order to solve this recurrence relation, we require two independent boundary conditions.
Once found, relation~\eqref{eq:rec1} yields the full steady-state solutions for all moments $I_n$, $B_n$, and $R_n$.
The first condition, related to the normalization or completeness relation, reads
\begin{align}
	I_{0} = \mean{\hat{1}} = 1.
\end{align}
The other one, usually applied by truncating the endless hierarchy of coupled equations, would follow from the boundedness of the density operator $\hat\varrho$. As $\lim\limits_{n\to\infty}\langle n|\hat\varrho|n\rangle=0$, we may set $\langle n_0|\hat\varrho|n_0\rangle=0$, thus cutting off the coupling between the different moments for $n<n_0$ and $n>n_0$, yielding a closed system of equations. The error from this approach can be adjusted by increasing $n_0$. The resulting boundary condition reads as
\begin{align}
  \langle n_0|\hat\varrho|n_0\rangle=&\frac{1}{n_0!}\left\langle:\hat a^{\dagger n_0}\hat a^{n_0}\exp(-\hat a^\dagger\hat a):\right\rangle\\
  =&\frac{1}{n_0!}\sum\limits_{k=0}^\infty\frac{(-1)^k}{k!}I_{n_0+k}=0,
\end{align}
where $\mean{:\cdot:}$ denotes normal ordering. In general, this  criterion is difficult to apply.
	Another possible formulation of the second condition is found using the fact that for any order $n$, the moment $I_{n}$ is positive semidefinite.
	Therefore, using Eq.~\eqref{eq:rec1}, we find
	\begin{align}
	0 \leq \alpha_{n+1} I_{n+1} + \beta_{n} I_{n}.
	\end{align}
	Since $\beta_{n} > 0$ holds for all $n$, this is equivalent to
	\begin{align}
	\frac{I_{n}}{I_{n+1}} \geq \frac{-\alpha_{n+1}}{\beta_{n}}.\label{eq:upperbound1}
	\end{align}
	For $\alpha_{n+1} \geq 0$, this inequality is trivial to fulfill. However, for  $\alpha_{n+1} < 0$, it yields a bound for the growth of the moments $I_{n}$ as
	\begin{align}\label{eq:upperbound}
	\frac{I_{n+1}}{I_{n}} \leq \frac{\beta_{n}}{-\alpha_{n+1}}.
	\end{align}
	Now, for sufficiently large $n$ we may only take into account the leading terms of the coefficients, which read (see Appendix~\ref{sec.DerAsymp} for details)
\begin{align}
   \alpha_{n+1}&\approx-\frac{\kappa^2}{4g^2}n^2<0,\label{eq.alpha_appr}\\
   \beta_n&\approx\frac{p}{2\kappa}n.\label{eq.beta_appr}
\end{align}
After inserting these into Eq.~\eqref{eq:upperbound}, we obtain
\begin{equation}\label{eq.asymprec}
  \frac{I_{n+1}}{I_n}\leq\frac{2g^2p}{\kappa^3}\frac{1}{n}=\xi\frac{1}{n},
\end{equation}
with the positive constant $\xi=2g^2p/\kappa^3$. As this ratio goes to zero, we can set as a second boundary condition 
\begin{equation}\label{eq:2ndBC}
 \lim\limits_{n\to\infty}I_n=0.
\end{equation}

A few things should be noted. First, Eq.~\eqref{eq:2ndBC} is by no means a general result for quantum states of bosonic systems. Quite the opposite, the moments for both a coherent state ($I_n^\text{coh}=I_1^n$) with $I_1>1$ and a thermal state ($I_n^\text{therm}=n!I_1^n$) with $I_1>0$ diverge. In particular, the latter is relevant as it would be the state in either of the (not physically reasonable) cases of $\kappa\to0$ or $p\to\infty$ (see Appendix~\ref{sec.p-asymp}). The other way around, we directly conclude that neither of these states is reached in this setup. Second, in the Appendix~\ref{sec.DerAsymp} we also estimated a lower bound of the ratio $I_{n+1}/I_{n}$, approaching again $\xi/n$ for large $n$. Hence, the right-hand side of Eq.~(\ref{eq.asymprec}) is not just an upper bound, but a very good approximation for the large-$n$ behavior. We will use this fact when we consider the characteristic function of the intracavity field. Third, as a consequence of our second note, we observe that neither $\delta$ nor $\Gamma$ play a role in the asymptotic behavior.  

The boundary condition~\eqref{eq:2ndBC} can be utilized to calculate a more practicable condition for $I_{1}$.
Iteratively applying the identity~\eqref{eq:rec1}, we obtain
\begin{align}
\label{eq:rec2}
I_{n+2} = C_{n+2} I_1 + D_{n+2} I_0.
\end{align}
The coefficients $C_{n+2}$ and $D_{n+2}$ obey recurrence relations like~\eqref{eq:rec1}, namely
\begin{align*}
C_{n+2} &= \alpha_{n+1} C_{n+1} + \beta_n C_n,\\
D_{n+2} &= \alpha_{n+1} D_{n+1} + \beta_n D_n,
\end{align*}
with initial conditions
\begin{align}
\label{eq:ICCD}
C_0 &= 0, \qquad C_1=1,\notag \\
D_0 &= 1, \qquad D_1 = 0.
\end{align}
	Applying the conditions $I_{0}=1$ and $ \lim\limits_{n\to\infty}I_n=0$, we find
\begin{align}\label{eq:I1}
I_1 = \lim_{n \to \infty}  - \frac{D_{n}}{C_{n}}.
\end{align}
	The calculation of the  moments $B_{n}$ and $R_{n}$ is straightforward using the steady-state form of Eq.~\eqref{eq:DES1_sub}.
	It should be kept in mind that for numerical calculations, a truncation has to be done in Eq.~\eqref{eq:I1}, where we have to set $I_{N}=0$ for an appropriate order $N$ and only moments up to this order can be calculated from Eq.~\eqref{eq:rec1}.
	
	Let us consider the numerical complexity of our method. Since the initial conditions~\eqref{eq:ICCD} are known and the $\alpha_{n+1}$ and $\beta_{n}$ are analytical functions in $n$, the recurrence relations for the $C_{N}$ and $D_{N}$ are of the order of~$\mathcal{O}(N)$. Using the recurrence relation~\eqref{eq:rec1} for the $I_{N}$ together with the calculated $I_{1}$, our approach yields the same complexity of $\mathcal{O}(N)$. Let us compare this to the standard approach for solving the steady state problem; cf.~\cite{Rice,PG2013}. The main step involves the inversion of the matrix of coefficients following directly from the master equation~\eqref{eq:master} governing the dynamic of the density matrix. Truncation at $N$ means a maximum of $N$ photons in the cavity, and thus $2(N+1)$ possible states. For the density matrix, this yields $[2(N+1)]^2 \propto N^2$ rows. A matrix inversion based on the standard Gauss-Jordan elimination algorithm scales in computational effort with~$\mathcal{O}(M^3)$, $M$ being the number of rows. Thus we obtain for a photon number cutoff of $N$ an~$\mathcal{O}(N^6)$ complexity scaling. In contrast to this, our method reduces the complexity to~$\mathcal{O}(N)$.

	We can also use the explicit formulas~\eqref{eq:upperbound} and \eqref{eq.asymprec} as well as the lower bound from the appendix, Eq.~\eqref{eq:upperlowerbound}, to infer an appropriate cutoff for the numerical calculations together with the corresponding error bars. First, one derives the bounds for $n$, for which the dependences in Eqs.~\eqref{eq.alpha_appr},~\eqref{eq.beta_appr} become dominant. Then we use the upper and lower bounds to determine the value for $n$, for which $I_{n+1}/I_{n}$ is within the desired $\varepsilon$ neighborhood. Applying Eq.~\eqref{eq:rec2} to Eq.~\eqref{eq:upperlowerbound}, with the assumption of $C_{n+1}>0$, we find
	\begin{equation}
	 -\frac{D_{n+1}-\frac{\beta_n}{\varepsilon-\alpha_{n+1}}D_n}{C_{n+1}-\frac{\beta_n}{\varepsilon-\alpha_{n+1}}C_n}\leq I_1\leq-\frac{D_{n+1}-\frac{\beta_n}{-\alpha_{n+1}}D_n}{C_{n+1}-\frac{\beta_n}{-\alpha_{n+1}}C_n}.\label{eq:error1}
	\end{equation}
	Comparing with Eq.~\eqref{eq:I1} and the approximation in Eqs.~\eqref{eq.alpha_appr} and~\eqref{eq.beta_appr}, we see that the correction to the applied formula Eq.~\eqref{eq:I1} is of the order $\mathcal{O}(n^{-3})$.

\begin{figure}[h]
	\includegraphics[width=8.2cm]{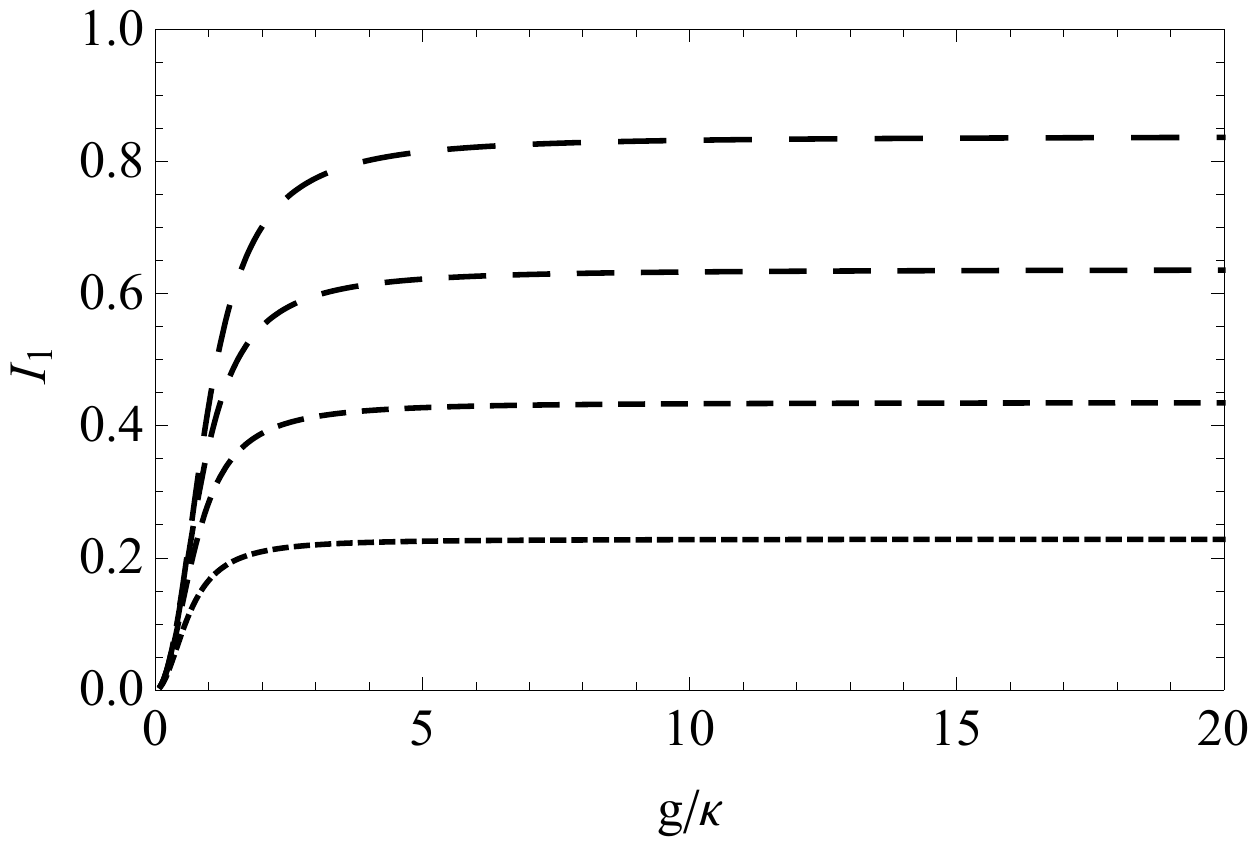}
	\caption{$I_{1}=\mean{\hat{a}^{\dagger}\hat{a}}$ for resonance $\delta=0$ as a function of $g/\kappa$.
	The normalized quantum-dot decay is $\Gamma/\kappa=1$ and the normalized pumping strength is varied from $p/\kappa =0.5$ to $2$ in steps of $0.5$, where a longer dashing corresponds to a stronger pumping.}
	\label{fig:intensity}
	\end{figure}

	As an example, we plotted in Fig.~\ref{fig:intensity} the first moment $I_{1}$, which is the intracavity field intensity, against the normalized coupling strength $g/\kappa$. The parameters are the same as in the previous work of Agarwal and Dutta Gupta (cf.~\cite{AD90}) and our method yields the same result.
	As the coupling increases, so does the intensity, and $I_{1}$ ultimately reaches a saturated state for large coupling. The saturation values appear to be linearly dependent on the pumping strength.
			
\begin{figure}[h]
		\includegraphics[width=8.2cm]{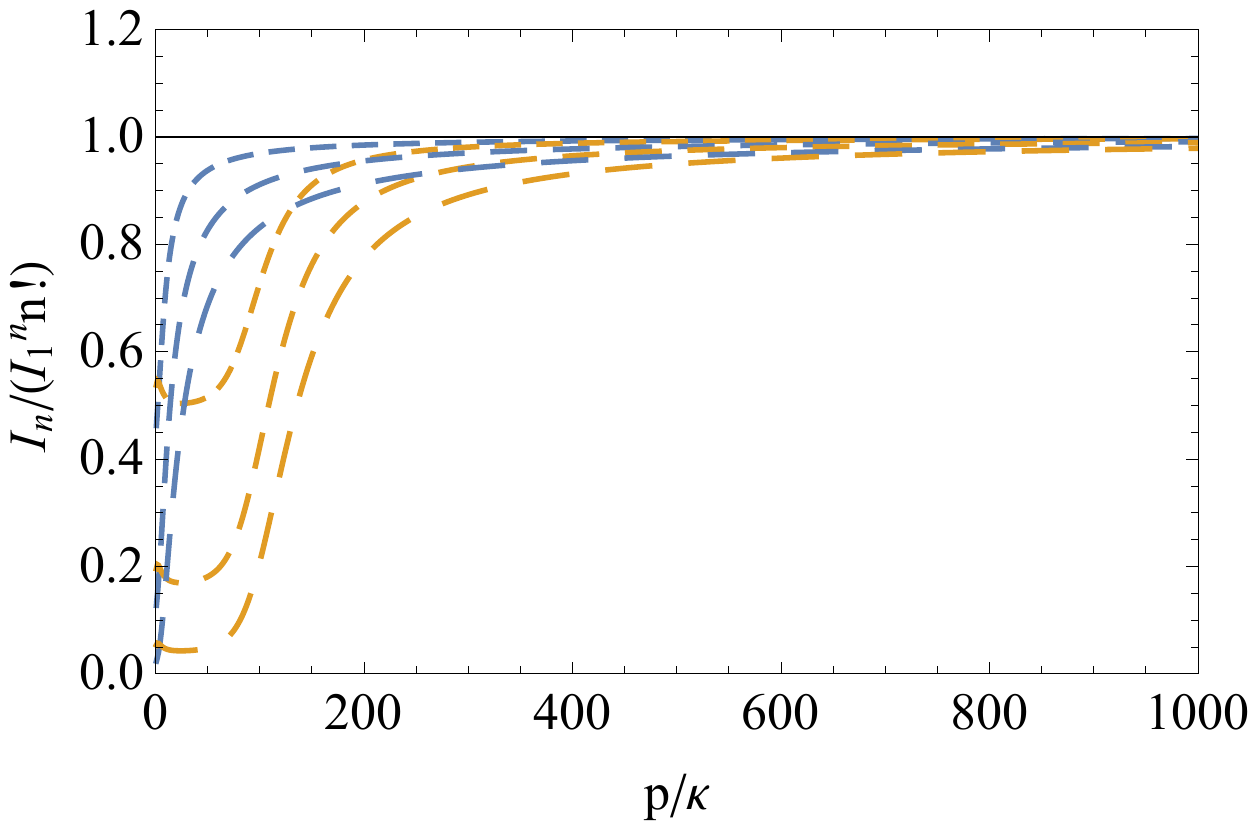}
		\caption{(Color online) Renormalized correlation functions $I_{n}/(I_{1}^{n} \, n!)$ for parameters $\Gamma/\kappa=1$, $\delta=0$, and $g/\kappa=1$ (blue/darker) or $g/\kappa=5$ (orange/lighter). Depicted are the orders $n=2,3,4$ with increasing dashing size, respectively.}
		\label{fig:corrn}
	\end{figure}

	Another example for the use of the moments is the calculation of the $n$th-order normalized intensity correlation of the intracavity field for zero time delay\citep{VW06}
	\begin{align}\label{eq:corrn}
	g^{(n)}(0) = \frac{I_{n}}{I_{1}^n}.
	\end{align}
	Those normalized correlations are, in principle, measurable by higher-order generalization of the antibunching experiment by Kimble \textit{et al}.~\cite{KDM77}, e.g., by time-multiplexing techniques~\cite{ALCS10}, and allow a comparison to well known states.
	For example a coherent state and a thermal state, as mentioned above, have moments $I_{n}^{\text{coh}}=I_{1}^{n}$ and  $I_{n}^{\text{therm}}=n! I_{1}^{n}$. Hence, their $n$th-order normalized correlations are equal to unity and $n!$, respectively.
	The intracavity field transits to a thermal state for $p\to\infty$, as can be seen from Fig.~\ref{fig:corrn}, where we plotted $I_{n}/(I_{1}^{n} \, n!)$ for $n=2,3,4$ against the normalized pumping strength. We chose two parameter sets with normalized coupling $g/\kappa=1$ (blue/darker) and $g/\kappa=5$ (orange/lighter), each with $\Gamma/\kappa=1$ and $\delta=0$. The curves tend for both sets and all orders to unity for increasing $p/\kappa$.
	It should be kept in mind that albeit the system converges formally into a thermal state for $p\to\infty$, it never reaches it. 
	The transition is only possible since in Eq.~\eqref{eq.asymprec} the upper bound~$\xi$ diverges, allowing the moments~$I_n$ to diverge. A rigorous proof of the thermal state limit is given in the Appendix~\ref{sec.p-asymp}.

\section{Moment based nonclassicality criteria}\label{sec:nonclass}

	We proceed with an investigation of nonclassicality based on the moments extracted from the steady-state.
	Nonclassical light fields are quantum fields with properties which cannot be described by the classic Maxwell equations and are thus interesting for many applications. Agarwal and Dutta Gupta applied their method~\cite{AD90} to analyze the Mandel $Q$ parameter~\cite{Ma79} of the intracavity field,
	\begin{equation}
	  Q=\frac{\langle(\Delta\hat n)^2\rangle-\langle\hat n\rangle}{\langle\hat n\rangle}=\frac{I_2-I_1^2-I_1}{I_1}.
	\end{equation}
	Determining this parameter in experiments requires a measurement of the photon statistics as the amplitude of mean value and variance of the photon number need to be compared. In the following, we will focus on the more general correlation conditions according to~\cite{AgTa,SRV05,EvgenNC05}. There, we only compare moments with the same powers of field operators. Hence, their verification in experiments can be directly concluded from field correlations measured in homodyne correlation~\cite{ShVoMeas06} setups.
	
	The following moment based criteria are calculated for two exemplary microcavity systems, namely, micropillars and microdisks. Those were realized in the last decade and give a good account of the realizable parameter ranges.
	The used parameters are taken from Khitrova \textit{et al}.~\cite{KGKKS06} and the two assigned parameter sets include the decay rates $\Gamma$ and $\kappa$ of the quantum dot (TLS) and cavity, respectively, as well as the coupling strength $g$.
	Note that according to the used Hamiltonian~\eqref{eq:hamilton}, all parameters are given as angular frequencies.
	The first set for the micropillars consists of $\Gamma = \SI{113e9}{\per\second}$, $\kappa=\SI{276e9}{\per\second}$, and $g=\SI{122e9}{\per\second}$ and is denoted as set~A.
	The second set, set~B, for the microdisks is $\Gamma = \SI{427e9}{\per\second}$, $\kappa=\SI{213e9}{\per\second}$, and $g=\SI{616e9}{\per\second}$.
	The remaining parameters, the pumping strength $p$ and detuning $\delta$, can then be subject to the experiment performed on those microcavities, thus being the independent variables.
	However, we will set $\delta=0$ since for our analysis the detuning is of less interest and we only vary $p$. 
	
	Experimentally accessible criteria for nonclassicality can be formulated by considering principle minors of the matrix of moments for suitable basic operators; cf.~\cite{AgTa,SRV05,EvgenNC05}.
	In our case, we define
	\begin{align}
	  \hat{f}=\sum\limits_{\substack{n\in\mathbb N\\k, l \in \lbrace 0 ,1\rbrace}} C_{n,k,l} \, \hat{a}^{\dagger \, n+l} \hat{a}^{n+k} \hat{A}_{21}^{k} \hat{A}_{12}^{l}.
	\end{align}
	Hence, the resulting matrix of moments is solely based on the moments defined in Eq.~\eqref{eq:momentsdef}.
	Then the necessary and sufficient condition for a nonclassical state is that at least one principle minor of $\langle:\hat f^\dagger \hat f:\rangle$ becomes negative. However, we will only analyze certain minors related to particular nonclassical effects, and thus the resulting conditions are just sufficient.
	The first condition deals with the intracavity field only and reads
	\begin{align}\label{eq:Incrit}
		\frac{I_{2n}}{I_{n}^2} < 1.
	\end{align}
	The special case of $n=1$ gives the well-known second-order correlation function for zero time delay, $g^{(2)}(0)$, and indicates sub-Poissonian photon statistics if $g^{(2)}(0) = I_{2}/I_{1}^2 <1$.

\begin{figure}[h]
	\includegraphics[width=8.2cm]{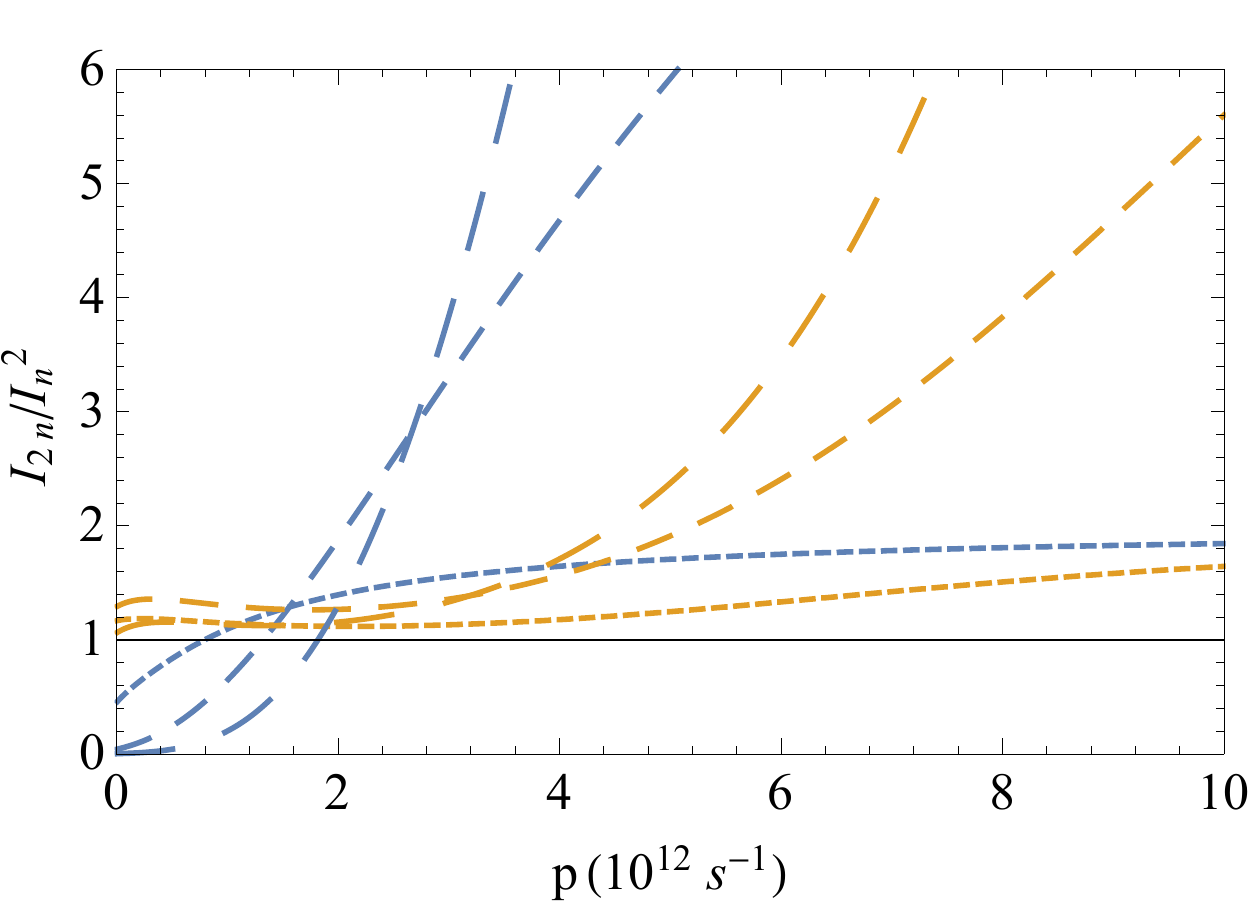}
	\caption{(Color online) Condition $I_{2n}/I_{n}^{2}$ plotted for orders $n=1,3,5$ and parameter sets~A~(blue/darker) and~B~(orange/lighter) for varied pumping strength $p$. A longer dashing corresponds to higher orders~$n$. Note that the order of magnitude of $p$ is $10^{3}$ times higher compared to the other parameters; see beginning of Sec.~\ref{sec:nonclass}.}
	\label{fig:Incrit}
	\end{figure}

	In Fig.~\ref{fig:Incrit}, we plotted the condition for orders $n=1,3,5$ against the pumping strength, in each case for both parameter sets~A and~B.
	For all shown orders $n$, set~A fulfills the condition~\eqref{eq:Incrit} below certain pumping strengths. Hence, the system undergoes a transition from the nonclassical to the classical regime for increasing pumping strength. Keeping the special case $n=1$ in mind, this would be the transition from a sub- to super-Poissonian photon statistics. The set~B does not fulfill the condition~\eqref{eq:Incrit} for any shown order~$n$. Note that this does not rule out any nonclassicalities of higher orders. In fact, set~B does fulfill the condition~\ref{eq:Incrit} for $n=10$.

\begin{figure}
	\includegraphics[width=8.2cm]{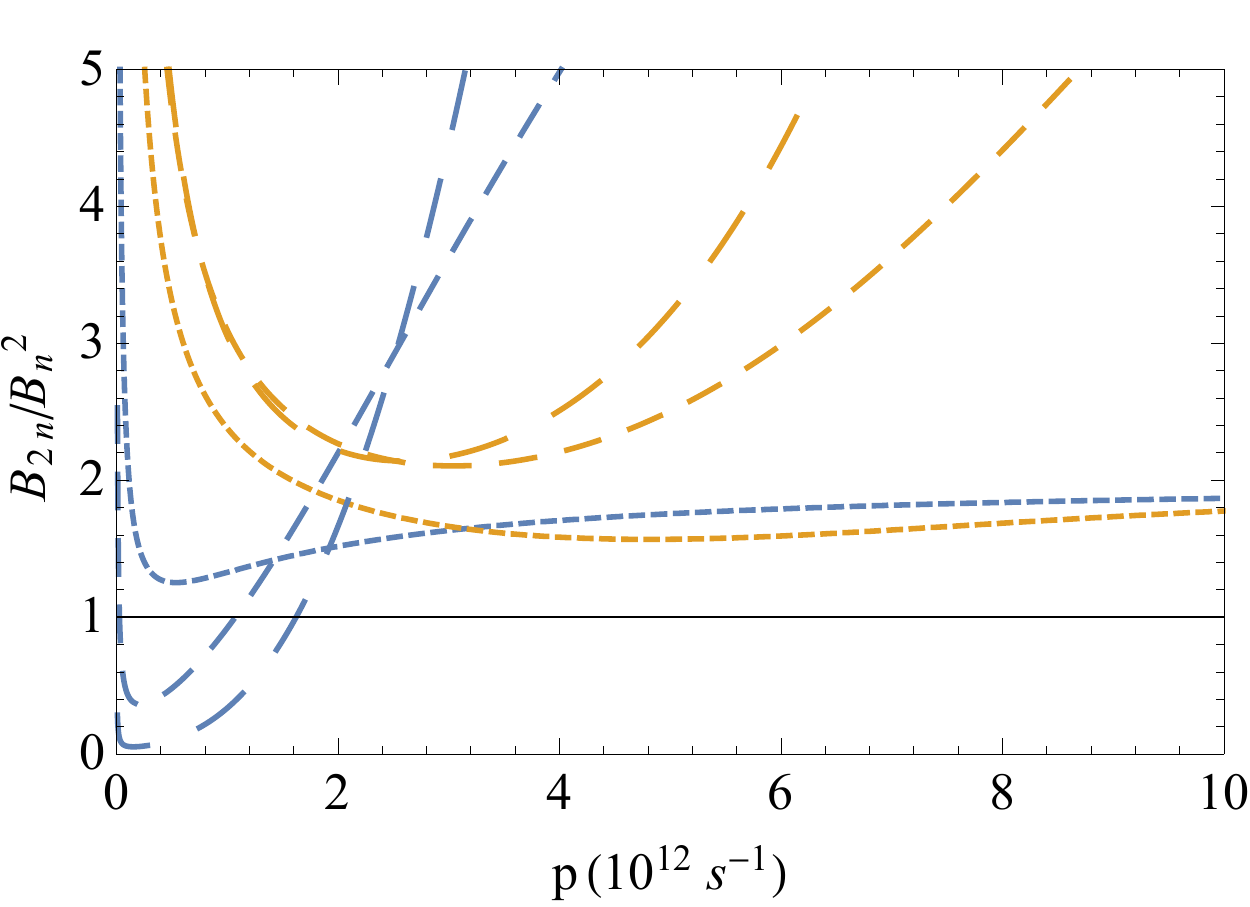}
	\caption{(Color online) $B_{2n}/B_{n}^{2}$ plotted for $n=1,3,5$ and parameter sets~A~(blue/darker) and B~(orange/lighter) against the pumping strength~$p$. A longer dashing corresponds to higher orders~$n$. Note that the order of magnitude of $p$ is $10^{3}$ times higher compared to the other parameters.}
	\label{fig:Bncrit}
	\end{figure}

	The second condition reads
	\begin{align}\label{eq:Bncrit}
		\frac{B_{2n}}{B_{n}^{2}} < 1,
	\end{align}
	which is formally akin to condition~\eqref{eq:Incrit} but takes into account the intensity of both the TLS and the intracavity field.
	Condition~\eqref{eq:Bncrit} is plotted against the pumping strength~$p$ in Fig.~\ref{fig:Bncrit} again for the sets~A~(blue/dark) and~B~(orange/light) and orders $n=1,3,5$. 
	For $n=1$, the condition is not fulfilled for both sets~A and~B, but higher-order criteria are fulfilled for set~A, again with consecutively lower values and for greater ranges of the pumping strength.
	This case again shows that the derived nonclassicality criteria are only sufficient and an unfulfilled criteria for some order $n$ does not rule out nonclassical effects of higher orders.
	
\begin{figure}
	\includegraphics[width=8.2cm]{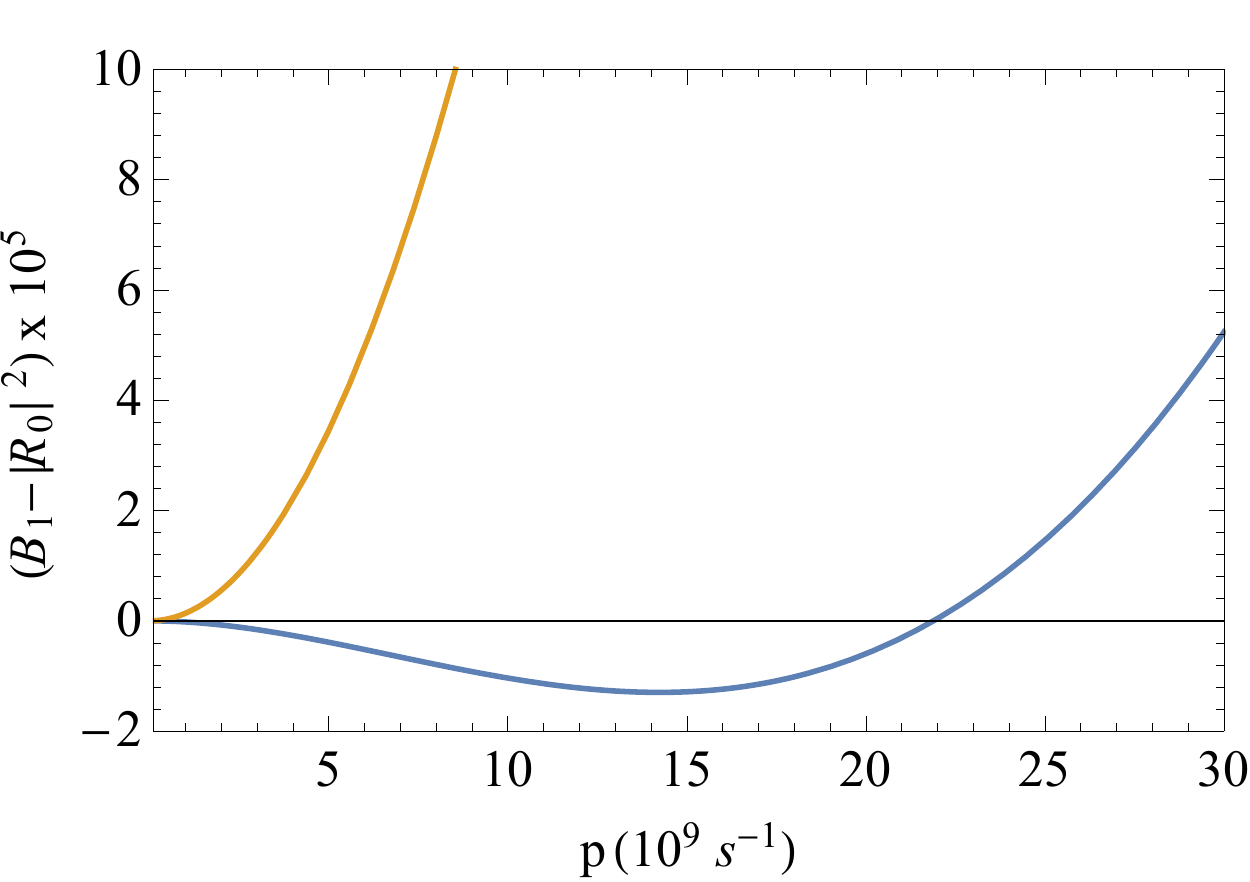}
	\caption{(Color online) Plot of $B_{1} - \left| R_{0} \right|^2$ for set~A~(blue/darker) and set~B~(orange/lighter). Negative values occur for entanglement between the intracavity field and the TLS.}
	\label{fig:entanglement}
	\end{figure}

	The third and last criterion examined here is
	\begin{align}\label{eq:entanglement}
	B_{1}/\left| R_{0} \right|^2 < 1.
	\end{align}
	While the more general relation
	\begin{align}
	B_{2n+1}/\left| R_{n} \right|^2 < 1
	\end{align}
	indicates nonclassicality as the other criteria, Eq.~(\ref{eq:entanglement}) is also a sufficient criterion for entanglement between the intracavity field and the TLS~\cite{ShVo05}. In Fig.~\ref{fig:entanglement}, we plotted condition~\eqref{eq:entanglement} for our two example systems.
	Similar to the case of condition~\eqref{eq:Incrit}, set~B does not show entanglement, whereas set~A fulfills the condition for certain pumping strengths.
	However, note the scaling of both the abscissa and ordinate.
	Negative values in Eq.~\eqref{eq:entanglement} only occur for very low pumping strengths and also attain only small values.
	
	The results so far indicate that set~A shows nonclassical behavior of different kinds, which reveal themselves via lower-order moment-based criteria. For set~B, on the other hand, those same order criteria do not indicate a nonclassical character. In particular, it should be noted that an accurate measurement of higher-order field moments, such as $I_{10}$ as shown in Fig.~\ref{fig:Incrit}, is currently out of technological reach~\cite{ALCS10}. This result is rather surprising as set~B has a better quantum-dot--cavity coupling relative to the dissipation rates. We conclude that the creation of nonclassical light in incoherently driven quantum-dot--cavity systems may be favored by not too strong coupling. In semiconductor quantum optics, strong coupling, and especially Rabi splitting, is considered a clear indicator of quantum light; see, in particular the argumentations in~\cite{QDstrong1,QDstrong2,QDstrong4}. Our results, on the other hand, clearly show the occurrence of strong signatures of nonclassicality for moderate coupling, making this scenario favorable for applications. 
	
	Detecting nonclassicality can be cumbersome, as there is an infinite hierarchy of sufficient but not necessary nonclassicality conditions~\cite{EvgenNC05}. For some system parameters, lower-order moments do not reveal any nonclassical effect. Hence, in the following, we will investigate the characteristic function of the intracavity field, which includes the full information also on higher-order moments. It is important that the characteristic function can be directly sampled from experimental data~\cite{LS02}. Moreover, the characteristic function may uncover quantum effects more directly then moments criteria~\cite{Ki09}.

\section{Characteristic function}\label{sec:charfct}
	The characteristic function is one possible representation of the quantum state of light, thus containing all information on the state. The same information is contained in all the moments $I_{n}$ and, in this section, we derive their relation to the characteristic function of the intracavity field. Using this result and the large-$n$ behavior of the $I_{n}$, the asymptotic behavior of the characteristic function is found. With this we prove that the intracavity field is nonclassical for any physically reasonable set of system parameters.

	The characteristic function is defined as~\cite{VW06,G63,S63}
	\begin{align}\label{eq:chardef}
		\Phi(\alpha) = \mean{:\hat{D}(\alpha):}, \qquad \alpha \in \mathbb{C}
	\end{align}
	with the displacement operator $\hat{D}(\alpha)=\exp(\alpha \hat{a}^{\dagger}-\alpha^{*}\hat{a})$.
	The representation of $\Phi(\alpha)$ in terms of the $I_{n}$ is derived by evaluating Eq.~\eqref{eq:chardef}, while bearing in mind that the off-diagonal elements $\mean{\hat{a}^{\dagger\,k}\hat{a}^{l}}_{k \neq l}$ vanish in the long-time limit due to the incoherent nature of the pumping.
	The result reads
	\begin{align}\label{eq:charfct}
		\Phi(\alpha) = \sum\limits_{n=0}^{\infty} \frac{(-1)^{n}}{(n!)^{2}} |\alpha|^{2n} I_{n}
	\end{align}
	and is therefore phase independent, $\Phi(\alpha)=\Phi(|\alpha|)$.
	For an actual calculation, we truncate the infinite sum at $n=N$ in accordance to the truncation for the moments $I_{N}=0$.
	
\begin{figure}
	\includegraphics[width=8.2cm]{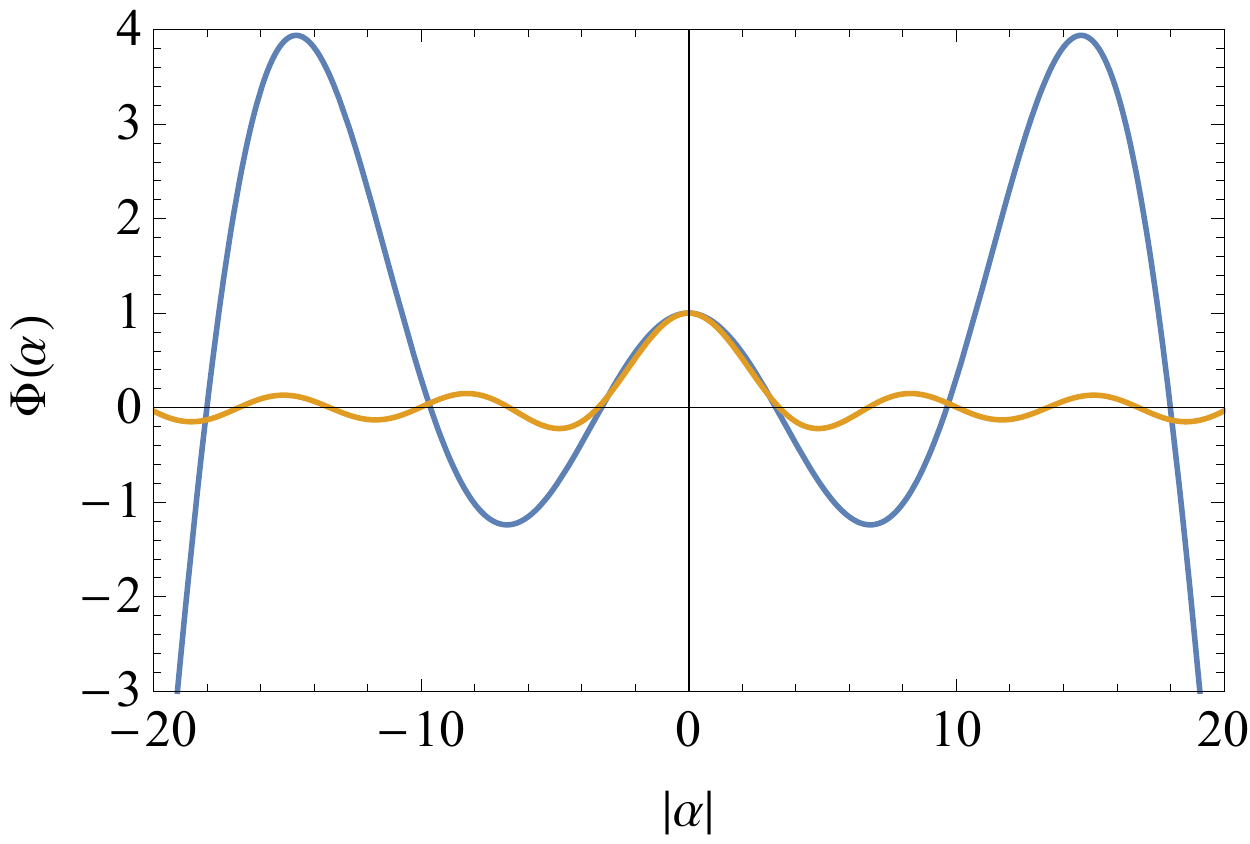}
	\caption{(Color online) Characteristic function for the parameter sets~A~(blue/darker) and~B~(orange/lighter) and pumping strength $p=\SI{1e11}{\per\second}$.}
	\label{fig:charfct}
	\end{figure}

	The characteristic function is of special interest since it contains the whole information about the underlying state and appropriate nonclassicality criteria can be formulated~\cite{RV02}.
	The simplest criterion reads~\cite{V00}
	\begin{align}\label{eq:charcrit}
		|\Phi(\alpha)| > 1.
	\end{align}
	In order to test Eq.~(\ref{eq:charcrit}), we plotted $\Phi(\alpha)$ in Fig.~\ref{fig:charfct} for the parameter sets~A and~B and a pumping strength of $p=\SI{1e11}{\per\second}$.
	In the plotted range of $|\alpha|$ only set~A fulfills the condition~\eqref{eq:charcrit} and therefore reveals nonclassicality.
	However, this analysis has its limits since only a restricted range of $|\alpha|$ can be displayed due to the truncation of the infinite sum in Eq.~\eqref{eq:charfct} at $n=N$.
	Beyond these limitations, we can, however, analyze the general behavior of the characteristic function when using the asymptotic behavior of the moments $I_{n}$.
	The detailed derivation can be found in Appendix~\ref{sec.DerAsymp}, which gives the asymptotic relation
	\begin{align}\label{eq:asymIn}
	I_{n+1}=\frac{\xi}{n} I_{n}, \qquad n\geq N,
	\end{align}
	with $\xi=2 p g^2 / \kappa^3$.
	Here, the order $N$ is defined as $N{:}= \text{max}(n_{1}, n_{2}, n_{3})$, where $n_{1}$, $n_{2}$, and $n_{3}$ are the orders for which the approximations $-\alpha_{n+1} \propto n^2$, $\beta_{n} \propto n$, and $I_{n+1}=I_{n} \, \beta_{n}/(-\alpha_{n+1})$ hold, respectively; cf. Eqs.~\eqref{eq:alphaseries}--~\eqref{eq:upperlowerbound}.
	
	It should be kept in mind that $N$ can become considerably large and for actual calculations like in Sec.~\ref{sec:nonclass} the relation~\eqref{eq:asymIn} is not reasonable.
	For a visualization, we plotted in Fig.~\ref{fig:upperlower} the upper and lower bounds for the moments according to Eq.~\eqref{eq:upperlowerbound} in Appendix~\ref{sec.DerAsymp} against the order $n$.
	We also added the asymptotic relation $\beta_{n}/(-\alpha_{n+1}) = \xi/n$ as used in Eq.~\eqref{eq:asymIn} and the case were the next-to-leading order is kept for the coefficients $\alpha_{n+1}$ and $\beta_{n}$; cf. Eqs.~\eqref{eq:alphaseries},\eqref{eq:betaseries}.
	The parameters were chosen from set~B with a pumping strength of $p=\SI{1e11}{\per\second}$.
	Apparently, the asymptotic relation holds true for orders $n\approx 10^3$, whereas the lower and upper bounds are already equal for $n\approx 10$. 
	Hence, the moments $I_{n}$ can already be approximated by the upper bound when the asymptotic relation still differs by orders of magnitude.

	\begin{figure}
	\includegraphics[width=8.2cm]{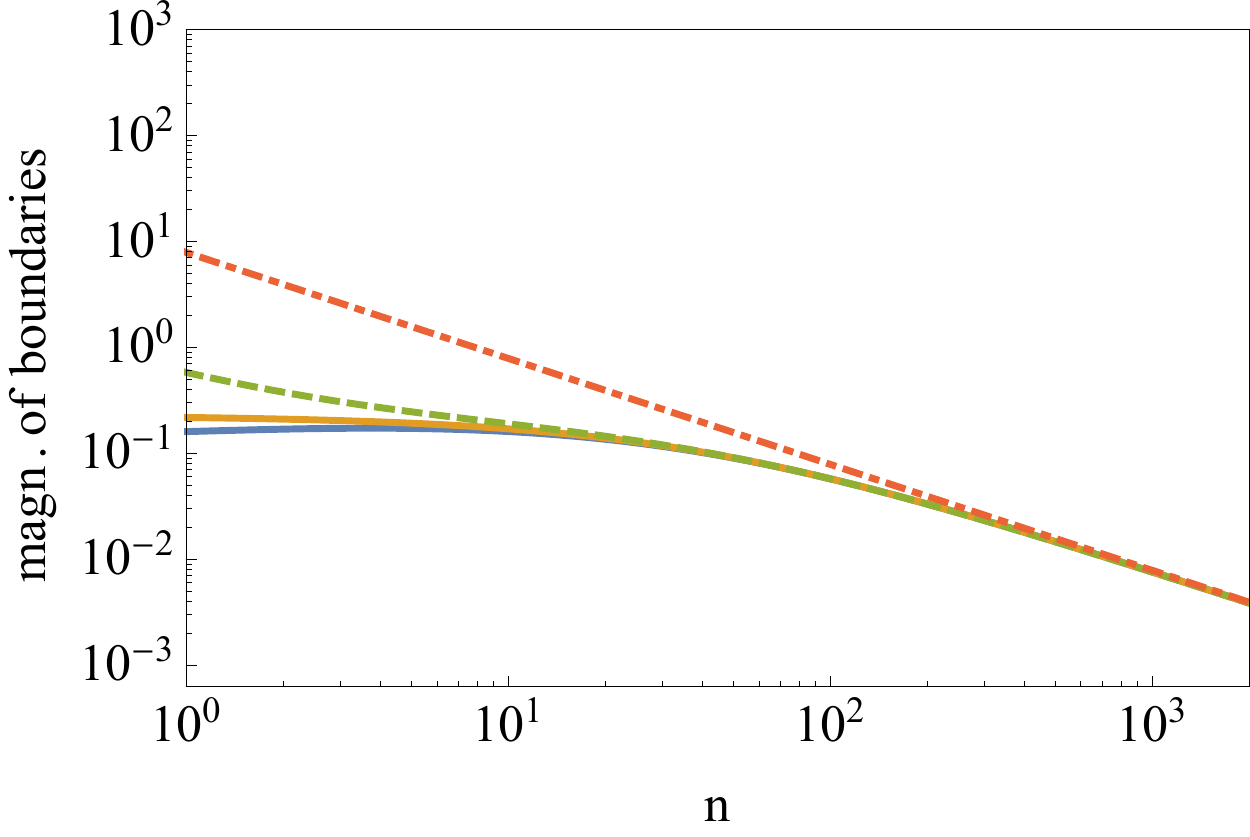}
	\caption{(Color online) Plot of the lower (blue/darker) and upper (orange/lighter) boundaries for the moments. Also seen are the approximated ratios $\beta_{n}/(-\alpha_{n+1})$ were the linear order is kept (green dashed line) and the asymptotic relations hold (red dot-dashed line). The parameters were taken from set~B and the pumping strength is $p=\SI{1e11}{\per\second}$.}
	\label{fig:upperlower}
	\end{figure}
	
	While the above considerations are important for actual calculations involving any of the mentioned approximations, we can neglect them for the following theoretical analysis and solely rely on the fact that there is some $N$ which fulfills the requirements of the asymptotic relation~\eqref{eq:asymIn}.
	Under these assumptions, we are now able to avoid the truncation but instead split the infinite sum in Eq.~\eqref{eq:charfct} into two parts. In the first sum, we use the exact moments up to the order $N$, and in the second, we insert the asymptotic relation~\eqref{eq:asymIn}.
	The latter sum can be simplified by iterative application of Eq.~\eqref{eq:asymIn}, leading to the following expression for $\Phi(\alpha)$
	\begin{align}\label{eq:charfctfull}
		\Phi(\alpha) &= \sum\limits_{n=0}^{N} \frac{(-1)^{n}}{(n!)^{2}} |\alpha|^{2n} I_{n} \notag \\&+ \frac{I_{N}(N-1)!}{\xi^{N}} \sum\limits_{n=N+1}^{\infty} \frac{(-1)^{n}}{(n!)^{3}} |\alpha|^{2n} n \xi^{n}.
	\end{align}
	With this expansion at hand, it can be shown that the asymptotic behavior of the characteristic function is sufficient to prove nonclassicality according to the criterion~(\ref{eq:charcrit}) for all physically reasonable sets of system parameters. At first we note that in general we do not know the coefficients of the first part of the expansion for the characteristic function as they follow from the full calculation. Hence, it is necessary for our aim to show that the asymptotic term increases faster than $|\alpha|^{2N}$ or any polynomial. Denoting $x=|\alpha|^2\xi$, the asymptotic behavior can be reduced to
	\begin{align}
		\Phi_\text{asymp}(x) &=V_N \sum\limits_{n=N+1}^{\infty}n \frac{(-x)^{n}}{(n!)^{3}},
	\end{align}
	with $V_N$ the positive prefactor. Note that $x>0$ allows to take positive roots of $x$.
	As we are only interested in the order of asymptotic behavior, the values of both the scaling $V_N$ and $N$ itself are irrelevant. The latter would only imply a subtraction of a polynomial of order $x^N$. We therefore set $V_N=1$ and $N=0$ to simplify our series to
	\begin{align}\label{eq:phiasymp}
		\Phi_\text{asymp}(x) &=\sum\limits_{n=1}^{\infty}n \frac{(-x)^{n}}{(n!)^{3}}.
	\end{align}
	Using Stirling's approximation, we obtain
	\begin{equation}
	  (n!)^3\approx\sqrt{(2\pi n)^3}\left(\frac{n}{e}\right)^{3n}\approx2\pi n3^{-(3n+1/2)}(3n)!
	\end{equation}
	and thus, after defining $\tilde x=\sqrt[3]{x}$
	\begin{align}\label{eq:n!to3n!}
		\sum\limits_{n=1}^{\infty}n \frac{(-x)^{n}}{(n!)^{3}}&\approx\frac{\sqrt{3}}{2\pi}\sum_{n=1}^\infty\frac{(-3\tilde x)^{3n}}{(3n)!}.
	\end{align}
	The sum on the right hand side of Eq.~(\ref{eq:n!to3n!}) can easily be obtained by a sum of three scaled exponential functions with complex exponents, yielding
	\begin{align}
	  \sum_{n=1}^\infty\frac{(-3\tilde x)^{3n}}{(3n)!}&=\frac{1}{3}\left[2\exp\left(\tfrac{3}{2}\tilde x\right)\cos\left(\tfrac{3\sqrt{3}}{2}\tilde x\right)+e^{-3\tilde x}\right]-1.
	\end{align}
	The addition of $-1$ stems from the limitation of the series to $n\geq1$. For $\tilde x\gg1$, both that constant term as well as the term $\exp(-3\tilde x)$ can be neglected. Inserting this result into Eq.~(\ref{eq:phiasymp}), we finally obtain
	\begin{align}\label{eq:phiasympfinal}
	  \Phi_\text{asymp}(x)&\approx \frac{1}{\sqrt{3}\pi}\exp\left(\tfrac{3}{2}x^{\tfrac{1}{3}}\right)\cos\left(\tfrac{3\sqrt{3}}{2}x^{\tfrac{1}{3}}\right)
	\end{align}
	This is an alternating, diverging function with an envelope function going with $\exp(\tfrac{3}{2}x^{\tfrac{1}{3}})$, which will overcome any polynomial increase or decrease from the first part of the characteristic function in Eq.~(\ref{eq:charfctfull}) or the approximation when setting $N=0$ in Eq.~\eqref{eq:phiasymp}. Hence, we can state as our final result that the steady-state intracavity field in our system is always nonclassical.

	\section{Summary}\label{sec:sum}
	We introduced an alternative technique for solving the steady-state problem of an incoherently pumped two-level system coupled to a single-mode cavity by using recurrence relations with appropriate boundary conditions. Due to the incoherent nature of the pumping, only three types of nonvanishing moments are needed, whose coupled equations of motion lead to the recurrence relation in the steady state. Boundary conditions have been derived and an analytical proof of convergence of the moments has been given. Furthermore, we provided a direct estimate of the errors in dependence on the cutoff procedure to be used in numerical calculations. Together with the method of solution via recurrence relations a significant reduction of numerical effort was achieved.
	
	Nonclassicality criteria based on moments have been analyzed for system parameters describing realistic microcavity structures. Nonclassical effects, such as sub-Poisson photon statistics for the intracavity field or its entanglement with the two-level system, occurred for certain parameters.
	The favorable scenario for demonstrating quantum phenomena is that of moderate quantum-dot--cavity coupling and weak pumping. This is contrary to the often discussed claim that strong coupling is essential for revealing nonclassical signatures. In the limit of extremely strong pumping, the quantum state of the cavity field approaches a thermal one. For more realistic conditions, however, the state is far from showing a thermal statistics.
	
	We also examined the characteristic function of the intracavity field, which contains the full information about the quantum state of light, based on the asymptotic relations describing the behavior of moments up to higher orders.
	This approach renders it possible to show that the intracavity field is nonclassical for any realistic set of parameters.
	In conclusion, the techniques developed in our paper provide helpful tools for describing a two-level system in a cavity in the steady state. They offer a variety of possibilities for characterizing nonclassicality of the intracavity field and its entanglement with the degrees of freedom of the two-level system.

\section*{acknowledgement} The authors gratefully acknowledge support by the Deutsche Forschungsgemeinschaft through SFB 652.

\appendix

\section{Lower bound and asymptotic behavior}\label{sec.DerAsymp}
	In addition to the upper bound for the ratio $I_{n+1}/I_{n}$ in Eq.~\eqref{eq:upperbound} we find a lower bound being valid from a certain order on.
	The recurrence relation~\eqref{eq:rec1} together with Eq.~\eqref{eq:upperbound} give
	\begin{align}
		I_{n+2} = \alpha_{n+1} I_{n+1} + \beta_{n} I_{n} \leq I_{n+1} \frac{\beta_{n+1}}{-\alpha_{n+2}}.
	\end{align}
	Hence, if there is a constant $\varepsilon \geq \beta_{n+1}/(-\alpha_{n+2}) > 0$, we have
	\begin{align}\label{eq:convergence}
		I_{n+2} = \alpha_{n+1} I_{n+1} + \beta_{n} I_{n} \leq \varepsilon I_{n+1}.
	\end{align}
	From the asymptotic behavior of the coefficients $\alpha_{n+1}$, $\beta_{n}$, which follow from their definitions~(\ref{eq:alpha} and~\ref{eq:beta}), and read
	\begin{align}\label{eq:alphaseries}
   \alpha_{n+1}&\approx-\frac{\kappa^2}{4g^2}n^2-\left[1+\frac{\kappa^2}{4g^2}(\tfrac{5}{2}+\tfrac{3}{2}\tfrac{\Gamma+p}{\kappa})\right]n,\\\label{eq:betaseries}
   \beta_n&\approx\frac{p}{2\kappa}(n+2),
\end{align}
	it follows that such a constant $\varepsilon$ does not only exist but that for every $\varepsilon > 0 $ there is an order $N$ from which Eq.~\eqref{eq:convergence} holds.
	Note that for the special case of $\varepsilon<1$, the moments $I_{n}$ converge monotonically.
	$N$ may be extracted from $\varepsilon \geq \beta_{n+1}/(-\alpha_{n+2})$ or more easily by using only the leading terms in Eqs.~(\ref{eq:alphaseries} and~\ref{eq:betaseries}) and $\xi=2p g^2/\kappa^3$ as in the result~(\ref{eq.asymprec}), giving $N\geq \xi/\varepsilon$.

	Now, we can also formulate a lower bound and together
	\begin{equation}\label{eq:upperlowerbound}
  		\frac{\beta_n}{\varepsilon-\alpha_{n+1}}\leq\frac{I_{n+1}}{I_n}\leq\frac{\beta_n}{-\alpha_{n+1}},\quad n\geq \frac{\xi}{\varepsilon}.
	\end{equation}
	For sufficiently large $n$, the relation between $I_n$ and $I_{n+1}$ is well approximated by $\xi/n$.

\section{Asymptotic behavior for large pumping}\label{sec.p-asymp}
In case of very large pumping strengths $p\gg\Gamma,\kappa,g,\delta$, we can simplify the coefficients $\alpha_{n+1}$ and $\beta_n$ to
\begin{align}
  \alpha_{n+1}\approx&\frac{p}{2\kappa}-\frac{p^2}{4g^2}-\frac{n+1}{2},\\
  \beta_n\approx&\frac{1+n}{2}\frac{p}{\kappa}.
\end{align}
Applying again our upper-bound scenario from Eq.~(\ref{eq:upperbound1}), we obtain
\begin{align}
  \frac{I_n}{I_{n+1}}\geq&\frac{\kappa}{p}-\frac{1}{n+1}+\frac{\kappa p}{2g^2(n+1)}\approx\frac{\kappa p}{2g^2(n+1)}.
\end{align}
The latter approximation follows again for sufficiently large pumping. As in the limit $p\to\infty$, the right-hand side can only be fulfilled for $I_{n+1}\to0$, we can approximate the behavior, similar to the case above, by
\begin{equation}
  I_{n+1}\approx\left(\frac{2g^2}{\kappa p}\right)(n+1)I_n.
\end{equation}
Repeatedly applying this formula and taking into account that for $p\to\infty$, the value of $n$ from which point on the asymptotic behavior is valid decreases, we find the relation of moments for a thermal state:
\begin{equation}
  I_n\approx n!\left(\frac{2g^2}{\kappa p}\right)^n.
\end{equation}
Consequently, the intensity $I_1$ is then given by $\tfrac{2g^2}{\kappa p}$.
In the strict limit of $p \to \infty$ the moments $I_{n}=0$, $\forall n \in \mathbb{N}$, and one obtains the vacuum state.


\begin{thebibliography}{99}
	\bibitem{P46} E. M. Purcell, Spontaneous emission probabilities at radio frequencies, Phys. Rev. \textbf{69}, 681 (1946).
	
    \bibitem{Rempe92} R. J. Thompson, G. Rempe and H. J. Kimble, Observation of normal-mode splitting for an atom in an optical cavity, Phys. Rev. Lett. \textbf{68}, 1132 (1992).
    
	\bibitem{Mollow} B. R. Mollow, Power spectrum of light scattered by two-level systems, Phys. Rev. \textbf{188}, 1969 (1969).
	
	\bibitem{Stroud} F. Schuda, C. R. Stroud, and M. Hercher, Observation of the resonant Stark effect at optical frequencies, J. Phys. B: At. Mol. Phys. \textbf{7}, L198 (1974).
	
	\bibitem{walther} F. Diedrich and H. Walther, Nonclassical Radiation of a Single Stored Ion, Phys. Rev. Lett. \textbf{58}, 203 (1987).
	
	\bibitem{toschek} M. Schubert, I. Siemers, R. Blatt, W. Neuhauser, and P. E. Toschek, Photon Antibunching and Non-Poissonian Fluorescence of a Single Three-Level Ion, Phys. Rev. Lett. \textbf{68}, 3016 (1992).
	
	\bibitem{C99} H. J. Carmichael, \textit{Statistical Methods in Quantum Optics 1} (Springer-Verlag Berlin Heidelberg, 1999).
	
	\bibitem{C08} H. J. Carmichael, \textit{Statistical Methods in Quantum Optics 2} (Springer-Verlag Berlin Heidelberg, 2008).
	
	\bibitem{ReiFor10} S. Reitzenstein and A. Forchel, Quantum dot micropillars, J. Phys. D: Appl. Phys. \textbf{43}, 033001 (2010).
	
	
	\bibitem{MichlerExp} K. Sebald, P. Michler, T. Passow, D. Hommel, G. Bacher, and A. Forchel, Single-photon emission of CdSe quantum dots at temperatures up to 200 K, Appl. Phys. Lett. \textbf{81}, 2920 (2002).
	
	\bibitem{Shih-2} E. B. Flagg, A. Muller, J. W. Robertson, S. Founta D. G. Deppe, M. Xiao, W. Ma, G. J. Salamo, and C. K. Shih, Resonantly driven coherent oscillations in a solid-state quantum emitter, Nat. Phys. \textbf{5}, 203 (2009).
	

	\bibitem{Shih-1} A. Muller, E. B. Flagg, P. Bianucci, X. Y. Wang, D. G. Deppe, W. Ma, J. Zhang, G. J. Salamo, M. Xiao, and C. K. Shih, Resonance Fluorescence from a Coherently Driven Semiconductor Quantum Dot in a Cavity, Phys. Rev. Lett. \textbf{99}, 187402 (2007).
	
	\bibitem{QDstrong1} J. P. Reithmaier, G. S\k{e}k, A. L\"offler, C. Hofmann, S. Kuhn, S. Reitzenstein, L. V. Keldysh, V. D. Kulakovskii, T. L. Reinecke, and A. Forchel, Strong coupling in a single quantum dot-semiconductor microcavity system, Nature (London) \textbf{432}, 197 (2004).
	
	\bibitem{QDstrong2} T. Yoshie, A. Scherer, J. Hendrickson, G. Khitrova, H. M. Gibbs, G. Rupper, C. Ell, O. B. Shchekin, and D. G. Deppe, Vacuum Rabi splitting with a single quantum dot in a photonic crystal nanocavity, Nature (London) \textbf{432}, 200 (2004).
	
	\bibitem{QDstrong4} J. Kasprzak, S. Reitzenstein, E. A. Muljarov, C. Kistner, C. Schneider, M. Strauss, S. H\"ofling, A. Forchel, and W. Langbein, Up on the Jaynes-Cummings ladder of a quantum-dot/microcavity system, Nat. Mat. \textbf{9}, 304 (2010).
	
	\bibitem{Xuan11} R.-W. Xuan, J.-P. Xu, X.-S. Zhang, P. Li, C.-Y. Luo, Y.-Y. Wu, and L. Li, Continuously voltage-tunable electroluminescence from a monolayer of ZnS quantum dots, Appl. Phys. Lett. \textbf{98}, 041907 (2011).
	
	\bibitem{QDstrong3} F. P. Laussy, E. Valle, und C. Tejedor, Strong Coupling of Quantum Dots in Microcavities, Phys. Rev. Lett. \textbf{101}, 083601 (2008).
	
	\bibitem{AD90} G. S. Agarwal, S. Dutta Gupta, Steady states in cavity QED due to incoherent pumping, Phys. Rev. A \textbf{42}, 1737 (1990).
	
	\bibitem{Ma79} L. Mandel, Sub-Poissonian photon statistics in resonance fluorescence, Opt. Lett. 4, 205 (1979).
	
	\bibitem{AgTa} G. S. Agarwal and K. Tara, Nonclassical character of states exhibiting no squeezing or sub-Poissonian statistics, Phys. Rev. A 46, 485 (1992).
	
	\bibitem{SRV05} E. Shchukin, Th. Richter, and W. Vogel, Nonclassicality criteria in terms of moments, Phys. Rev. A \textbf{71}, 011802(R) (2005).
	
	\bibitem{EvgenNC05} E. Shchukin and W. Vogel, Nonclassicality moments and their measurements, Phys. Rev. A \textbf{72}, 043808 (2005).

	\bibitem{V00} W. Vogel, \text{Nonclassical States: An Observable Criterion}, Phys. Rev. Lett. \textbf{84}, 1849 (2000).		
	
        \bibitem{RV02} Th. Richter and W. Vogel, Nonclassicality of Quantum States: a Hierarchy of Observable Conditions, Phys. Rev. Lett. \textbf{89}, 283601 (2002).
	
	\bibitem{JC63} E. Jaynes and F. Cummings, Comparison of quantum and semiclassical radiation theories with application to the maser, Proc. IEEE \textbf{51}, 89 (1963).
	
	\bibitem{Rice} J. P. Clemens, P. R. Rice, P. K. Rungta, and R. J. Brecha, Two-level atom in an optical parametric oscillator: Spectra of transmitted and fluorescent fields in the weak-driving-field limit, Phys. Rev. A \textbf{62}, 033802 (2000).
	
	\bibitem{PG2013} P. Gr\"unwald and W. Vogel, Optimal squeezing in the resonance fluorescence of single-photon emitters, Phys. Rev. A \textbf{88}, 023837 (2013).
	
	\bibitem{VW06} W. Vogel and D.-G. Welsch, {\it Quantum Optics}, 3rd ed. (Wiley-VCH, Weinheim, 2006).
	
	\bibitem{KDM77} H. J. Kimble, M. Dagenais, and L. Mandel, Photon Antibunching in Resonance Fluorescence, Phys. Rev. Lett. \textbf{39}, 691 (1977)
	
	\bibitem{ALCS10} M. Avenhaus, K. Laiho, M. V. Chekhova, and C. Silberhorn, Accessing Higher Order Correlations in Quantum Optical States by Time Multiplexing, Phys. Rev. Lett. \textbf{104}, 063602 (2010).
	
	\bibitem{ShVoMeas06} E. Shchukin and W. Vogel, Universal Measurement of Quantum Correlations of Radiation, Phys. Rev. Lett. \textbf{96}, 200403 (2006).
	
	\bibitem{KGKKS06} G. Khitrova, H. M. Gibbs, M. Kira, S. W. Koch, and A. Scherrer, Vacuum Rabi splitting in semiconductors, Nature Phys. \textbf{2}, 81 (2006).
	
	\bibitem{ShVo05} E. Shchukin and W. Vogel, Inseparability Criteria for Continuous Bipartite Quantum States, Phys. Rev. Lett. \textbf{95}, 230502 (2005).
	
	\bibitem{LS02} A. I. Lvovsky and J. H. Shapiro, Nonclassical character of statistical mixtures of the single-photon and vacuum optical states, Phys. Rev. A \textbf{65}, 033830 (2002). 
	
	\bibitem{Ki09} T. Kiesel, W. Vogel, B. Hage, J. DiGuglielmo, A. Samblowski, and R. Schnabel, Experimental test of nonclassicality criteria, Phys. Rev. A \textbf{79}, 022122 (2009).
	
	\bibitem{G63} R. J. Glauber, Coherent and incoherent states of the radiation field, Phys. Rev. \textbf{131}, 2766 (1963).
	
	\bibitem{S63} E. C. G. Sudarshan, Equivalence of Semiclassical and Quantum Mechanical Description of Statistical Light Beams, Phys. Rev. Lett. \textbf{10}, 277 (1963).
	
\end{thebibliography}
\end{document}